\documentclass[sigconf]{acmart}
 \usepackage{textcomp, libertine}
\usepackage{booktabs} 
\usepackage[normalem]{ulem}
\usepackage{multirow}
\usepackage{bm}
\useunder{\uline}{\ul}{}
\usepackage{subcaption}
\usepackage{enumitem}
\usepackage[ruled]{algorithm2e}
\usepackage{amssymb}
\hypersetup{
,bookmarksnumbered=true%
,hypertexnames=false%
,breaklinks=true%
,colorlinks=true%
,linkcolor=black%
,citecolor=black%
,urlcolor=black%
}

\definecolor{tealblue}{rgb}{0.21, 0.46, 0.53}
\definecolor{wildstrawberry}{rgb}{1.0, 0.26, 0.64}
\definecolor{ao(english)}{rgb}{0.0, 0.5, 0.0}

\def\ignore#1{}

\begin{document}
\title{Learning to Selectively Transfer: Reinforced Transfer Learning for Deep Text Matching}

\author{Chen Qu}
\authornote{Work done during an internship at Alibaba Group.}
\affiliation{
	\institution{University of Massachusetts Amherst}
}
\email{chenqu@cs.umass.edu}

\author{Feng Ji, Minghui Qiu}
\authornote{Corresponding author}
\affiliation{
	\institution{Alibaba Group}
}
\email{{zhongxiu.jf, minghui.qmh}@alibaba-inc.com}

\author{Liu Yang}
\affiliation{
	\institution{University of Massachusetts Amherst}
}
\email{lyang@cs.umass.edu}

\author{Zhiyu Min}
\affiliation{
	\institution{Carnegie Mellon University}
}
\email{zhiyum@andrew.cmu.edu}

\author{Haiqing Chen, Jun Huang}
\affiliation{
	\institution{Alibaba Group}
}
\email{{haiqing.chenhq, huangjun.hj}@alibaba-inc.com}

\author{W. Bruce Croft}
\affiliation{
	\institution{University of Massachusetts Amherst}
}
\email{croft@cs.umass.edu}



\begin{abstract}
Deep text matching approaches have been widely studied for many applications including question answering and information retrieval systems.
To deal with a domain that has insufficient labeled data, these approaches can be used in a Transfer Learning (TL) setting to leverage labeled data from a resource-rich source domain.
To achieve better performance, source domain data selection is essential in this process to prevent the ``negative transfer" problem.
However, the emerging deep transfer models do not fit well with most existing data selection methods, because the data selection policy and the transfer learning model are not jointly trained, leading to sub-optimal training efficiency.

In this paper, we propose a novel reinforced data selector to select high-quality source domain data to help the TL model.
Specifically, the data selector ``acts" on the source domain data to find a subset for optimization of the TL model, and the performance of the TL model can provide ``rewards" in turn to update the selector.
We build the reinforced data selector based on the actor-critic framework and integrate it to a DNN based transfer learning model, resulting in a Reinforced Transfer Learning (RTL) method.
We perform a thorough experimental evaluation on two major tasks for text matching, namely, paraphrase identification and natural language inference.
Experimental results show the proposed RTL can significantly improve the performance of the TL model.
We further investigate different settings of states, rewards, and policy optimization methods to examine the robustness of our method.
Last, we conduct a case study on the selected data and find our method is able to select source domain data whose Wasserstein distance is close to the target domain data. This is reasonable and intuitive as such source domain data can provide more transferability power to the model.



\end{abstract}


\copyrightyear{2019}
\acmYear{2019}
\setcopyright{acmcopyright}
\acmConference[WSDM '19]{The Twelfth ACM International Conference on Web Search and Data Mining}{February 11--15, 2019}{Melbourne, VIC, Australia}
\acmBooktitle{The Twelfth ACM International Conference on Web Search and Data Mining (WSDM '19), February 11--15, 2019, Melbourne, VIC, Australia}
\acmPrice{15.00}
\acmDOI{10.1145/3289600.3290978}
\acmISBN{978-1-4503-5940-5/19/02}
\fancyhead{}
\settopmatter{printacmref=true, printfolios=false}

\maketitle

\section{Introduction}
\label{sec:intro}

Text matching is an important problem in both information retrieval and natural language processing. Typical examples of text matching include paraphrase identification~\cite{Socher2011PoolRAEPI}, natural language inference~\cite{SNLI}, document retrieval~\cite{Guo2016ADR}, question answering (QA)~\cite{Yang2016aNMMRS}, and conversational response ranking~\cite{Yang2018RespRank}. In particular, text matching plays a key role in conversational assistant systems to answer customer questions automatically. For example, the Contact Center AI\footnote{https://cloud.google.com/solutions/contact-center/} recently launched by Google and the AliMe~\cite{Li2017AliMe} built by Alibaba Group are both capable of handling informational requests by retrieving potential answers from a knowledge base. 

We illustrate the importance of text matching by describing the role it plays in a retrieval-based QA system. 
Typically, for a given user query, the system measures its similarity with the questions in the knowledge base and returns the answer of the best matched question~\cite{Yan2016LearningToRespond,Yu2018TL}.
The query-question matching problem can be modeled as a paraphrase identification (PI) or natural language inference (NLI) task, which are both typical tasks of text matching. Thus in this work, we focus on PI and NLI tasks to evaluate the performance of our method on text matching. We believe the improvement of text matching methods can benefit the end tasks such as question answering.
PI and NLI problems have been widely studied in previous work~\cite{SNLI, Yin2016ABCNN, Yin2018NLI, Socher2011PoolRAEPI}. However, when applied to real world applications, such methods face the challenge of insufficient labeled data in different domains. For example, in the E-commerce industry, a QA system has to handle each small domain of products such as books, electronics, clothes, etc. It is unrealistic to obtain a large amount of labeled training data for every small domain. As a promising approach to bridge the domain discrepancy, Transfer Learning (TL) has become an important research direction in the past several years~\cite{Yu2018TL, Shen2018WassersteinDG, ruder2017select, Yosinski2014TLVision, Liu2017TLNLP}.

Due to the domain shift between the source and target domains, directly applying TL approaches may result in ``negative transfer" problem. To prevent this problem, we argue that source domain data selection is necessary for the TL approaches. Table~\ref{tab:intro-example} gives an example of negative transfer in the PI task. ``Order" typically means to place an order for a product in the E-commerce domain (target domain). However, in an open domain (source domain) dataset, ``order" can be used to denote a succession or sequence. Hence in such case, TL without source domain data selection might result in negative transfer. 

\begin{table*}[htbp]
\footnotesize
\caption{An example of negative transfer in the PI task. This table
is best viewed in color. ``Order" in blue means to place an order for a product, which is typical in the E-commerce domain. ``Order" in red means a succession or sequence, which might appear in the source open domain. Transfer learning without source domain data selection might result in negative transfer.}
\label{tab:intro-example}
\vspace{-0.3cm}
\begin{tabular}{@{}lll@{}}
\toprule
Domain                  & Sentence 1                                                         & Sentence 2                                                                            \\ \midrule
                         & Which answers does Quora show first for each question?             & How does Quora decide the \textcolor{red}{\textbf{\textit{order}}} of the answers to a question?  \\
                         & What \textcolor{red}{\textbf{\textit{order}}} should the Matrix movies be watched in                  & Is there any particular \textcolor{red}{\textbf{\textit{order}}} in which I should watch the Madea movies                \\
\multirow{-3}{*}{\begin{tabular}[c]{@{}l@{}}Source\\(Open Domain)\end{tabular}}   & How can I \textcolor{blue}{\textbf{order}} a cake from Walmart online?                        & How do I \textcolor{blue}{\textbf{order}} a cake from Walmart?                                                   \\ \midrule
                         & How long is my \textcolor{blue}{\textbf{order}} arriving? Is it over? Will I have the refund? & I have escalated an \textcolor{blue}{\textbf{order}} and have not been updated in over a week.                   \\
                         & How can i get an \textcolor{blue}{\textbf{order}} receipt or invoice?                                  & How do I get an invoice to pay? \\

\multirow{-3}{*}{\begin{tabular}[c]{@{}l@{}}Target\\(E-commerce Domain)\end{tabular}} & I need to understand why my \textcolor{blue}{\textbf{orders}} have been cancelled             & Why my \textcolor{blue}{\textbf{order}} have been closed?                                                        \\ \bottomrule
\end{tabular}
\end{table*}

Recently, neural architectures are employed to leverage a large amount of source domain data and a small amount of target domain data in a multi-task learning manner~\cite{Mou2016TL,Yang2017TL}, which can be described as Deep Neural Networks (DNN) based supervised transfer learning. 
The DNN based TL framework has been proven to be effective in deep text matching tasks for question answering systems~\cite{Yu2018TL}.
Although various data selection methods~\cite{ruder2017select, Chen2011CoTrainingTLDataSelection, Huang2006CorrectSamplingBiasTL, Patel2018TLDataSelectionWithRL} were proposed for TL settings, most of them do not fit well with neural transfer models, because the data selector/reweighter is not jointly trained with the TL model. Specifically, the TL task model is considered as a sub-module of the data selection framework. Thus the TL task model needs to be retrained repetitively to provide sufficient updates to the data selection framework. Due to the relatively long training time of neural models, such data selection methods may suffer from long training time when applied to neural TL models. Therefore, we argue that data selection methods for transfer learning need to be revisited under the DNN based TL setting.

In the setting of DNN based transfer learning, the TL model is updated with mini-batch gradient descent in an iterative manner. In order to learn a universal data selection policy in this setting, we model the problem of source domain data selection as a Markov Decision Process (MDP)~\cite{Puterman1994MDP}. Specifically, at each time step (mini-batch/iteration), the TL model is at a certain state $s$, the decision maker (data selector) chooses an action $a$ to select samples from the current source batch to optimize the TL model. The TL model gives the data selector a reward $r$ and moves on to the next state $s'$. 
The state of $s'$ depends on the current state $s$ and the action $a$ made by the data selector. 
To solve this problem, it is intuitive to employ reinforcement learning, where the decision maker is the data selection policy that needs to be learned.

In this paper, we propose a novel reinforced data selector to select high-quality source data to help the TL model. Specifically, we build our data selector based on the actor-critic framework and integrate it to a DNN based TL model, resulting in a Reinforced Transfer Learning (RTL) method.
To improve the model training efficiency, the instance based decisions are made in a batch. Rewards are also generated on a batch level. 
Extensive experiments on PI and NLI tasks show that our RTL method significantly outperforms existing methods. 
Finally, we use Wasserstein distance to measure the target and source domain distances before and after the data selection. 
We find our method is able to select source data whose Wasserstein distance is close to the target domain data. This is reasonable and intuitive as such source domain data can provide more transferability power to the model.

Our contributions can be summarized as follows. (1) To the best of our knowledge, we propose the first reinforcement learning based data selector to select high-quality source data to help the DNN based TL model. (2) In contrast to conducting data selection instance by instance, we propose a batch based strategy to sample the actions in order to improve the model training efficiency. (3) We perform thorough experimental evaluation on PI and NLI tasks that involves four benchmark datasets. We find that the proposed reinforced data selector can effectively improve the performance of the TL model and outperform several existing baseline methods. We also use Wasserstein distance to interpret the model performance.

\section{Related Work}
\label{sec:relatedwork}




\textbf{Paraphrase Identification and Natural Language Inference.} PI and NLI problems have been extensively studied in previous work. Existing methods include using convolutional, recurrent, or recursive neural networks to model the sentence interactions, attentions, or encoding of a pair of input sentences~\cite{SNLI, Yin2016ABCNN, Yin2018NLI, Socher2011PoolRAEPI}. All the methods have been proven to be highly effective if given enough labeled training data. However, in real world applications, obtaining a large amount of labeled data by human annotation is not always affordable in terms of time and expense. Therefore, we focus on PI and NLI tasks in a transfer learning setting in this paper.

\textbf{Transfer Learning.} Transfer learning has been widely studied in the past years~\cite{Pan2010TLSurvey}. Existing work can be mainly classified into two categories. The first category makes the assumption that labeled data from both source and target domains are available to us, though the amount may differ~\cite{Daume2007TL, Yu2018TL}. While the second category assumes that no labeled data from the target domain is available in addition to the labeled source domain data~\cite{Shen2018WassersteinDG,ruder2017select}. Our work falls into the first category. In addition, an alternative view of taxonomy on transfer learning is to focus on methods. In this case, there are also two categories. The first is the instance based methods, which select or reweight the source domain training samples so that data from the source domain and the target domain would share a similar data distribution~\cite{ruder2017select, Chen2011CoTrainingTLDataSelection, Huang2006CorrectSamplingBiasTL}. The second category is feature based methods, which aim to locate a common feature space that can reduce the differences between the source and target domains. This goal is accomplished either by transform the features from one domain to be closer to the other domain, or to project both domains into a common latent space~\cite{Yu2018TL,Shen2018WassersteinDG}. 

In terms of instance based methods and feature based methods, our work falls into the first category, and we select data from the source domain to benefit the task performance in the target domain. In this line of work, data selectors/reweighters are typically not jointly trained with the TL model, which can lead to negative impacts on training efficiency~\cite{ruder2017select, Chen2011CoTrainingTLDataSelection, Patel2018TLDataSelectionWithRL}. Specifically, the TL model is considered as a sub-module of the data selection framework and the data selection policy is updated based on the final performance of the TL model. Due to the relatively long training time of neural models, such data selection methods suffer from poor training efficiency if applied to neural transfer learning models.

\textbf{Reinforcement Learning.} The concept of reinforcement learning (RL) dates back to several decades ago~\cite{DBLP:books/lib/SuttonB98,SurveyDRL}. Recent advances in deep learning made it possible for RL agents to generate dialogs~\cite{Li2016Dialog}, play video games~\cite{Mnih2013Atari}, and even outperform human experts in the game of Go~\cite{Silver2017MasteringTG}. Reinforcement learning algorithms can be categorized into two types: value based methods and policy based methods. Value based methods estimate the expected total return given a state. This type of method includes SARSA~\cite{Rummery1994Sarsa} and the Deep Q Network~\cite{Mnih2015DQN}. Policy based methods try to find a policy directly instead of maintaining a value function, such as the REINFORCE algorithm~\cite{DBLP:journals/ml/Williams92Reinforce}. Such methods can provide strong learning signals to update the policy. Finally, it is also possible to combine the value based and policy based approaches for a hybrid solution, such as the actor-critic algorithm~\cite{Konda2003ActorCritic}. This employs a learned value estimator to provide a baseline for the policy network for variance reduction. We experiment with policy based methods and hybrid methods in our model. 

Given the dynamic nature of reinforcement learning, researchers found it useful to employ RL in data selection problems, because data selection during training can be modeled as a sequential decision making process. So far, RL has been applied to data selection in active learning~\cite{Meng2017ActiveLearnRL}, co-training~\cite{Wu2018RLCoTraining}, and other applications of supervised learning, including computer vision~\cite{Fan2017Learn, Patel2018TLDataSelectionWithRL}, machine reading comprehension~\cite{Wang2018R3RR}, and entity relation classification~\cite{Feng2018RLEntity}. However, there is a lack of reinforced data selection methods under a DNN based transfer learning setting.
\section{Our Approach}
\label{sec:our-approach}

In this section, we present our reinforced transfer learning (RTL) framework under a DNN based transfer learning setting. The reinforced data selector is integrated into a TL model to select source domain data to prevent negative transfer. 

\subsection{Task Definition}
\label{subsec:task-definition}
We formulate our task into three subtasks: a text matching task, a transfer learning task, and a data selection task.


\subsubsection{\textbf{Text Matching.}}
Both paraphrase identification and natural language inference tasks can be unified as a text matching problem, defined as follows. Given two sentences $ \mathbf{X}_1 = \{ \mathbf{x}_1^1, \mathbf{x}_2^1, \dots, \mathbf{x}_m^1 \} $ and $ \mathbf{X}_2 = \{ \mathbf{x}_1^2, \mathbf{x}_2^2, \dots, \mathbf{x}_n^2 \} $, where $ \mathbf{x}_i^j $ denotes a word embedding vector either randomly initialized or retrieved from a pre-trained global vector look up table (such as GloVe~\cite{pennington2014glove}). $m$ and $n$ denotes the lengths of $\mathbf{X}_1$ and $\mathbf{X}_2$ respectively. The goal is to predict a binary label $y \in \{0, 1\}$ that denotes whether $\mathbf{X}_1$, $\mathbf{X}_2$ are semantically related. For PI, $y = 1$ denotes the two sentences are semantically identical (PARAPHRASE). For NLI\footnote{The common NLI task contains a third label of CONTRADICTION, which denotes the premise and the hypothesis are contradicted. The SciTail~\cite{scitail} dataset in our experiment does not come with this label.}, $y = 1$ denotes that the hypothesis $\mathbf{X}_2$ can be inferred from the premise $\mathbf{X}_1$ (ENTAILMENT). 

\subsubsection{\textbf{Transfer Learning.}} We consider the transductive transfer learning setting, where the source and target tasks are the same, while the source and target domains are different~\cite{Pan2010TLSurvey}. In contrast to conventional transductive TL where no labeled target domain data is available, we assume some target domain training data is available to perform supervised training of a base model. However, we expect a significantly larger amount of source domain training data can boost the performance of the aforementioned base model under a transfer learning setting. Given the labeled source domain data $\mathcal{D}_s$ and labeled target domain data $\mathcal{D}_t$ for the same task, where $|\mathcal{D}_s| >> |\mathcal{D}_t|$, the TL model leverages both $\mathcal{D}_s$ and $\mathcal{D}_t$ to improve the performance of the base model in the target domain.

\subsubsection{\textbf{Data Selection.}} The data selection task under a transfer learning setting is formulated as follows. A transfer learning algorithm updates the TL model with a batch of source data $\mathcal{X}_b^s$ and a batch of target data $\mathcal{X}_b^t$ iteratively. $\mathcal{X}_b^s$ and $\mathcal{X}_b^t$ are drawn from $\mathcal{D}_s$ and $\mathcal{D}_t$ respectively. The data selection module intervenes before the update of every iteration. Specifically, the data selection module selects a part of data $\mathcal{X}_b^{s'}$ from $\mathcal{X}_b^s$ according to a policy $\pi$. The selected $\mathcal{X}_b^{s'}$ is expected to produce a better performance than $\mathcal{X}_b^s$ after this single iteration as well as future iterations.
 
\subsection{Overview}
\label{subsec:overview}
\begin{figure}
    \centering
    \includegraphics[width=0.49\textwidth]{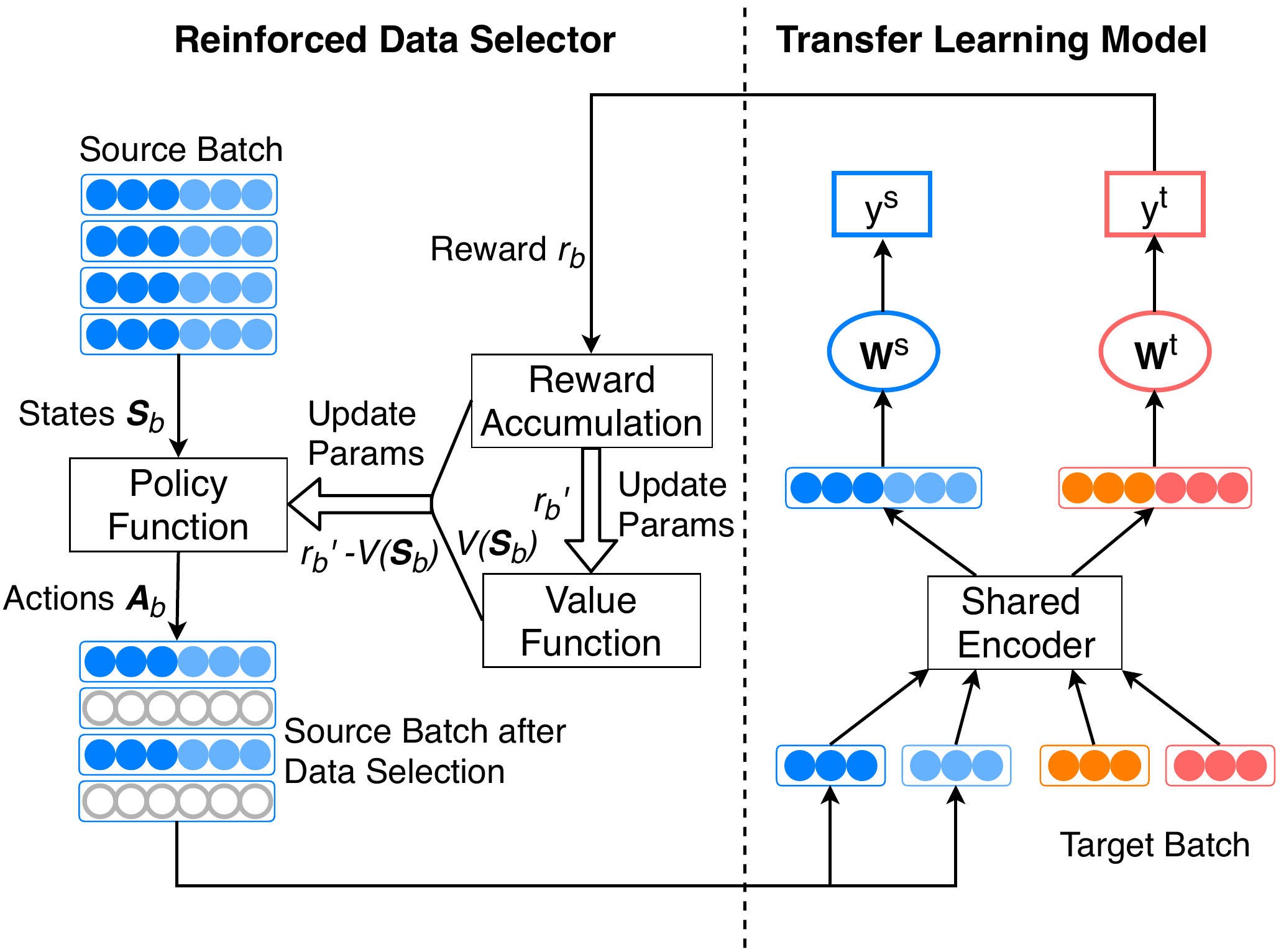}
    \vspace{-0.6cm}
    \caption{Architecture of the proposed RTL framework, which consists of two major parts: a reinforced data selector and a TL model. This figure is best viewed in color. The ``Shared Encoder" refers to the base model embedded in the TL model. The reinforced data selector selects a part of the source batch (blue) and feeds them into the TL model at each iteration. The TL model generates a reward on the target domain validation data for the data selector. Target batches (orange/pink) are fed into the TL model without data selection.}
    \label{fig:overview}
\end{figure}

The proposed framework consists of three components: a base model, a transfer learning model, and a reinforced data selector, corresponding to the above three subtasks respectively. The base model tackles the basic problem of text matching. It takes in a pair of sentences and generates a hidden representation of the sentence pair for the final prediction. The TL model is built on the top of the base model to leverage a large amount of source domain data. Finally, the reinforced data selector is a compartmentalized part in the transfer learning framework to handle the data selection for source domain data. The reinforced data selector is designed to prevent negative transfer and thus maximize the effectiveness of the TL model. Figure~\ref{fig:overview} gives an overview of our model.

\subsection{Base Model}
\label{subsec:base-model}
As illustrated in Figure~\ref{fig:overview}, the base model in our method is a shared encoder (shared neural network).
Note that our method is a general framework which can integrate different base models. 
Any implementation of text matching models can be adapted to our framework. 
However, for real-world applications, it is a common practice to consider the efficiency at both training and testing time~\cite{Yu2018TL}. Thus, we use the Decomposable Attention Model (DAM)~\cite{DBLP:conf/emnlp/Parikh2016DAM} as our base model, as DAM has effective performance with remarkable efficiency in text matching. 

DAM consists of three jointly trained components: ``attend", ``compare", and ``aggregate". First, the ``\textbf{attend}" module softly aligns the input pair of two sentences by obtaining the unnormalized attention weights $e_{i,j}$. Formally,
\begin{equation}\label{eqn:dam-attention-weights}
\begin{aligned}
e_{ij} = F(\mathbf{x}_i^1)^\top F(\mathbf{x}_j^2)
\end{aligned}
\end{equation}
where $F(\cdot)$ is a feed-forward network. Then the attention weights are normalized as follows:
\begin{equation}\label{eqn:dam-attention-weights-norm}
\begin{aligned}
\epsilon_i^2 = \sum_{j=1}^n \text{softmax}^j(e_{ij})\, \mathbf{x}_j^2, \quad
\epsilon_j^1 = \sum_{i=1}^m \text{softmax}^i(e_{ij})\, \mathbf{x}_i^1
\end{aligned}
\end{equation}
where $\text{softmax}^k(\cdot)$ is to perform the softmax function on dimension $k$. $\epsilon_i^2$ is interpreted as the subphrase in $\mathbf{X}_2$ aligned to $\mathbf{x}_i^1$.

Then the ``\textbf{compare}" module compares the aligned subphrases separately and produces a set of matching vectors as follows:
\begin{equation}\label{eqn:dam-compare}
\begin{aligned}
\mathbf{v}_{1,i} = G(\mathbf{x}_i^1 \oplus \epsilon_i^2), \quad
\mathbf{v}_{2,j} = G(\mathbf{x}_j^2 \oplus \epsilon_j^1)
\end{aligned}
\end{equation}
where $\oplus$ means concatenation and $G(\cdot)$ is a feed-forward network.

Finally, the ``\textbf{aggregate}" module combines the matching vectors to produce a representation of the sentence pair for a final prediction. Formally, the aggregated vectors are computed as follows:
\begin{equation}\label{eqn:dam-aggregate}
\begin{aligned}
\mathbf{v}_1 = \sum_{i=1}^m \mathbf{v}_{1,i},\quad
\mathbf{v}_2 = \sum_{j=1}^n \mathbf{v}_{2,j}. 
\end{aligned}
\end{equation}
Here $\mathbf{v}_1$ can be viewed as performing a sum pooling over the concatenated matrix $[\mathbf{v}_{1,1}, \mathbf{v}_{1,2}, \cdots, \mathbf{v}_{1,m} ]$. In practice, it's also beneficial to consider a max pooling over the matrix~\cite{Chen2017LSTMNLI}, i.e.:
\begin{equation}\label{eqn:dam-aggregate-1}
\begin{aligned}
\mathbf{v}_1^{max} = \text{max}_{i=1}^m (\mathbf{v}_{1,i}), \quad 
\mathbf{v}_2^{max} = \text{max}_{j=1}^n (\mathbf{v}_{2,j})
\end{aligned}
\end{equation}

The aggregated vectors are concatenated to form the output $\mathbf{z}$:
\begin{equation}\label{eqn:dam-output}
\begin{aligned}
\mathbf{z} = \mathbf{v}_1 \oplus \mathbf{v}_2 \oplus \mathbf{v}_1^{max} \oplus \mathbf{v}_2^{max}
\end{aligned}
\end{equation}

The base model can be formulated as a transformation function $f$ that takes in a pair of sentences $(\mathbf{X}_1(i), \mathbf{X}_2(i))$ as input and produces a hidden representation $\mathbf{z}_i = f(\mathbf{X}_1(i), \mathbf{X}_2(i))$. If the base model is used alone to make predictions, a classification (fully-connected) layer will be added after obtaining the hidden representation.

\subsection{Transfer Learning Model}
\label{subsec:transfer-learning-model}
As shown in Figure~\ref{fig:overview}, we consider a DNN based transfer learning framework with a fully-shared encoder~\cite{Mou2016TL,Yang2017TL}. The proposed data selection method is a general method that can be adapted to other TL frameworks, including fully-shared and specific-shared models~\cite{Yu2018TL}. We only consider the fully-shared model because we would like to keep the TL model simple and focus on the reinforced data selector. The base model serves as a shared encoder for sentence pairs from both source domain and target domain. For each given pair of sentences $(\mathbf{X}_1^d, \mathbf{X}_2^d)$ from domain $d$ ($d\in\{s, t\}$), the shared encoder maps $(\mathbf{X}_1^d, \mathbf{X}_2^d)$ to a hidden representation $\mathbf{z}^d = f(\mathbf{X}_1^d, \mathbf{X}_2^d)$. Then a classification layer maps $\mathbf{z}^d$ to a label $y^d$. The classification layers are separately learned for different domains. Formally,
\begin{equation}\label{eqn:clf_layer}
\begin{aligned}
p(y^d \mid \mathbf{z}^d) = 
\text{softmax}(\mathbf{W}^d \mathbf{z}^d+\mathbf{b}^d)
\end{aligned}
\end{equation}
where $\mathbf{W}^d$ and $\mathbf{b}^d$ are the weight matrix and bias vector for the classification layer respectively for domain $d$. Thus, the training objective is to minimize the training loss for both source and target domains. For training, we use the cross-entropy loss as follows.

\begin{equation}\label{eqn:loss_d}
\begin{aligned}
\mathcal{L}_d = \frac{1}{N^d} \sum_{n=1}^{N^d} - \text{log}\, p(y_c^{(n)} | \mathbf{X}_1^d(n), \mathbf{X}_2^d(n)), \quad d\in\{s, t\}
\end{aligned}
\end{equation}
where $y_c^{(n)}\in \{0,1\}$ is the ground truth label for the $n$-th data pair.

The fully-shared encoder and the source classification layer is considered as the \textit{source model}. Similarly, the fully-shared encoder and the target classification layer is considered as the \textit{target model}.

\subsection{Reinforced Data Selector}
\label{subsec:reinforced-data-selector}

\subsubsection{\textbf{Overview}}
\label{subsubsec:RL-overview}
We cast the source domain data selection in a transfer learning setting as a Markov Decision Process, which can be solved by reinforcement learning. The reinforced data selector is an agent that interacts with the environment constructed by the TL model. The agent takes actions of keeping or dropping a given source sample (a sentence pair) according to a learned policy. The agent bases its  decision on a state representation that describes several features of the given sample. The TL model evaluates the agent's actions and produces rewards to guide the learning of the agent. The agent's goal is to maximize the expected future total rewards it receives.

Reinforcement learning is commonly used for policy learning of agents in video games or chess games. In the context of video games, an episode commonly refers to a round of the game where the player either passes or fails the game at the end. The player would take a sequence of actions to reach the terminal state. In a neural transfer learning setting, the TL model is updated batch by batch for several epochs. It is natural to consider an epoch as an episode and a batch as a step to take actions.

As shown in Figure~\ref{fig:overview}, given each batch of the source domain sentence pairs $ \mathcal{X}_b^s = \{(\mathbf{X}_1(i), \mathbf{X}_2(i))_{i=1}^n\} $, where $b$ denotes the batch ID and $n$ denotes the batch size. We obtain a batch of states $\mathcal{S}_b = \{\mathbf{S}_1, \mathbf{S}_2, \dots, \mathbf{S}_n\}$, where $\mathbf{S}_i$ denotes the state of the $i$-th sentence pair $(\mathbf{X}_1(i), \mathbf{X}_2(i))$. Then the reinforced data selector makes a decision for each sample $(\mathbf{X}_1(i), \mathbf{X}_2(i))$ according to a learned policy $\pi(\mathbf{S}_i)$. Actions are also made in batch, denoted as $\mathcal{A}_b = \{a_1, a_2, \dots, a_n\}$, where $a_i \in \{0, 1\}$. $a_i = 0$ means to drop $(\mathbf{X}_1(i), \mathbf{X}_2(i))$ from $\mathcal{X}_b^s$. Thus, we obtain a new source domain batch $\mathcal{X}_b^{s'}$ that contains the selected source samples only. Finally, The transfer model is updated with $\mathcal{X}_b^{s'}$ and produce a reward $r_b$ according to the  performance on target domain validation data.

We will introduce the state, action, and reward in the following sections. The batch ID $b$ is omitted in some cases for simplicity.

\subsubsection{\textbf{State}}
\label{subsubsec:RL-state}
The state of a given source domain sentence pair $(\mathbf{X}_1(i), \mathbf{X}_2(i))$ is denoted as a continuous real valued vector $\mathbf{S}_i \in \mathbb{R}^l$, where $l$ is the dimension of the state vector. $\mathbf{S}_i$ represents the concatenation of the following features:
\begin{enumerate}[leftmargin=2em]
    \item A hidden representation $\mathbf{z}_i$, which is the output of the shared encoder given $(\mathbf{X}_1(i), \mathbf{X}_2(i))$.
    \item The training loss of $(\mathbf{X}_1(i), \mathbf{X}_2(i))$ on the  source model.
    \item The testing loss of $(\mathbf{X}_1(i), \mathbf{X}_2(i))$ on the target model.
    \item The predicted probabilities of $(\mathbf{X}_1(i), \mathbf{X}_2(i))$ on the source model.
    \item The predicted probabilities of $(\mathbf{X}_1(i), \mathbf{X}_2(i))$ on the target model.
\end{enumerate}
The first feature is designed to present the raw content to the data selector. Feature (3) and Feature (5) are based on the intuition that helpful source domain training data would be classified with relatively high confidence on the target model. Feature (2) and feature (4) are also provided as feature (3) and feature (5)'s counterparts on the source model. 

\subsubsection{\textbf{Action}}
\label{subsubsec:RL-action}
An action is denoted as $a_i \in \{0, 1\}$, which indicates whether to drop or keep $(\mathbf{X}_1(i), \mathbf{X}_2(i))$ from the source batch. $a_i$ is sampled according to a probability distribution produced by a learned policy function $\pi(\mathbf{S}_i)$. $\pi(\mathbf{S}_i)$ is approximated with a policy network that consists of two fully-connected layers. Formally, $\pi(\mathbf{S}_i)$ is defined as follows:
\begin{equation}\label{eqn:policy-1}
\begin{aligned}
\pi(\mathbf{S}_i) &= P(a_i|\mathbf{S}_i)
= \text{softmax}(\mathbf{W}_2 \mathbf{H}_i + \mathbf{b}_2) \\
\mathbf{H}_i &= \text{tanh}(\mathbf{W}_1 \mathbf{S}_i + \mathbf{b}_1)
\end{aligned}
\end{equation}
where $\mathbf{W}_k$ and $\mathbf{b}_k$ are the weight matrix and bias vector respectively for the $k$-th layer in the policy network, and $\mathbf{H}$ is an intermediate hidden state. 

\subsubsection{\textbf{Reward}}
\label{subsubsec:RL-reward}
The data selector takes actions to select data from $\mathcal{X}_b^s$ and form a new batch of source data $\mathcal{X}_b^{s'}$. We use $\mathcal{X}_b^{s'}$ to update the source model and obtain an immediate reward $r_b$ with a reward function $R(\mathcal{S}_b, \mathcal{A}_b)$. In contrast to conventional reinforcement learning, where one action is sampled based on one state and obtaining one reward from the environment, our actions are sampled in a batch based on a batch of states and obtaining one reward in order to improve model training efficiency. 

The reward is set to the prediction accuracy on the target domain validation data for each batch. Other metrics generated on the target validation data could also be applicable. To accurately evaluate the utility of $\mathcal{X}_b^{s'}$, $r_b$ is obtained after the source model is updated and before the target model is updated. For the extremely rare case of $\mathcal{X}_b^{s'} = \emptyset$, we skip the update of source model for this step. 

We compute the future total reward for each batch after an episode. Formally, $r_b'$ for batch $b$ is computed as follows.

\begin{equation}\label{eqn:discount-future-reward}
\begin{aligned}
r_b' = \sum_{k=0}^{N-b} \gamma^k r_{b+k}
\end{aligned}
\end{equation}
where $N$ is the number of batches in this episode, $r_b'$ is the future total reward for batch $b$, and $\gamma$ is the reward discount factor.

\subsubsection{\textbf{Optimization}}
\label{subsubsec:RL-optimization}
We experiment with two methods to update the policy network: the REINFORCE algorithm~\cite{DBLP:journals/ml/Williams92Reinforce} and the actor-critic algorithm~\cite{Konda2003ActorCritic}. 
Our model is mainly based on the actor-critic framework since it can help to reduce the variance so that the learning is more stable~\cite{Konda2003ActorCritic}. 

For any given episode, we aim to maximize the expected total reward. Formally, we define the objective function as follows:
\begin{equation}\label{eqn:objective}
\begin{aligned}
J(\Theta) = E_{\pi_{\Theta}}[\sum_{b=1}^n r_b]
\end{aligned}
\end{equation}
where the policy function $\pi$ is parameterized by $\Theta$. We further compute the gradient to make a step of update as follows:
\begin{equation}\label{eqn:gradient}
\begin{aligned}
\Theta \leftarrow \Theta + \alpha \frac{1}{n} \sum_{i=1}^n v_i \nabla_\Theta \text{log}\, \pi_\Theta(\mathbf{S}_i)
\end{aligned}
\end{equation}
where $\alpha$ is the learning rate, $n$ is the batch size, and $v_i$ is the target that guides the update of the policy network. The actor-critic algorithm combines policy based methods and value based methods for stable updates. We employ a value estimator network as the value function to estimate the future total reward $V_\Omega(\mathbf{S}_i)$ for each state $\mathbf{S}_i$ in a given batch. Thus, $v_i$ is computed as follows:
\begin{equation}\label{eqn:actor-critic-target}
\begin{aligned}
v_i = r_b' - V_\Omega(\mathbf{S}_i)
\end{aligned}
\end{equation}

The structure of the value estimator network is similar to the policy network except that the output layer is a regression function. Formally, the value network is optimized to approximate the real future total reward $r_b'$, i.e. to minimize the Mean Squared Error (MSE) between $V_\Omega(\mathbf{S}_i)$ and $r_b'$:
\begin{equation}\label{eqn:value-gradient}
\begin{aligned}
\Omega \leftarrow \Omega + \alpha \frac{1}{n} \sum_{i=1}^n \nabla_\Omega \text{MSE}(r_b', V_\Omega(\mathbf{S}_i))
\end{aligned}
\end{equation}
where the value function $V$ is parameterized by $\Omega$. 

In addition to the actor-critic algorithm described above, we also experiment with a Policy Gradient method named REINFORCE algorithm. In this case, $v_i$ is simply set to $r_b'$ in Equation~\ref{eqn:gradient}, which means that every action $a_i$ in this batch shares the same reward $r_b'$.

\subsection{Model Training}
\label{subsec:model-training}
The TL model and the reinforced data selector are learned jointly as they interact with each other closely during training. To optimize the policy network, we use the actor-critic algorithm described in Section~\ref{subsubsec:RL-optimization}. To optimize the TL model, we use a gradient descent method to minimize the loss function in Equation~\ref{eqn:loss_d}. We first pre-train the TL model for $k$ iterations and then start the joint training process. We use such a procedure following previous work~\cite{Feng2018RLEntity, Bahdanau2017ActorCritic}.

The details of the joint learning process is described in Algorithm~\ref{alg:joint-learning}. When optimizing the TL model, the gradient is computed based on a batch of training data. The TL model leverages training data in both source and target domains for better model performance. The reinforced data selector intervenes before every iteration of source model update by selecting helpful source domain data. Therefore, the intervention process has an impact on the gradient computed for the source model update, which includes the update for the shared encoder. The TL model provides a reward in turn to evaluate the utility of the data selection. After each epoch/episode, the policy network is updated with the actor-critic algorithm with the stored (states, actions, reward) triplets.

\begin{algorithm}[ht]
\footnotesize
\SetKwInOut{Input}{Input}

\SetAlgoLined
\Input{Episode $L$, source domain training data $\mathcal{D}_s$, target domain training data $\mathcal{D}_t$ and validation data $\mathcal{D}_t^{val}$}
 Initialize the pre-trained source and target model in TL model\;
 Initialize the policy network and value estimator network\;
 \For{episode $l = 1$ to $L$}{
  Obtain the random batch sequence: $\mathcal{D}_s = \{\mathcal{X}_1^s, \mathcal{X}_2^s, \dots, \mathcal{X}_N^s\}$ and $\mathcal{D}_t = \{\mathcal{X}_1^t, \mathcal{X}_2^t, \dots, \mathcal{X}_N^t\}$\;
  
  \ForEach{$(\mathcal{X}_b^s, \mathcal{X}_b^t)$ in $\{(\mathcal{X}_b^s, \mathcal{X}_b^t)_{b=1}^N\}$}{
    
    Obtain the states $\mathcal{S}_b = \{\mathbf{S}_1, \mathbf{S}_2, \dots, \mathbf{S}_n\}$ for $\mathcal{X}_b^s$\;
    
    Sample actions $\mathcal{A}_b$ according to the policy $\pi(\mathcal{S}_b)$\;
    
    Obtain the filtered source training batch $\mathcal{X}_{b}^{s'}$\;
    
    Update the source model with $\mathcal{X}_b^{s'}$\;
    
    Obtain the reward $r_b$ on the target model with $\mathcal{D}_t^{val}$\;
    
    Update the target model with $\mathcal{X}_b^t$\;
    
    Store $(\mathcal{S}_b, \mathcal{A}_b, r_b)$ to an episode history $H$\;
    
  }
  
  \ForEach{$(\mathcal{S}_b, \mathcal{A}_b, r_b)$ in $H$}{
    
    Obtain the future total reward $r_b'$ as in Eq.~\ref{eqn:discount-future-reward}\;
    
    Obtain the estimated future total rewards $V(\mathcal{S}_b)$\;
    
    Update the policy network following Eq.~\ref{eqn:gradient}\;
  }
  
  \ForEach{$(\mathcal{S}_b, \mathcal{A}_b, r_b)$ in $H$}{
    
    Obtain the future total reward $r_b'$ as in Eq.~\ref{eqn:discount-future-reward}\;
    
    Update the value estimator network as in Eq.~\ref{eqn:value-gradient}\;
  }
  
  Empty $H$\;
 }
 \caption{Joint learning of the transfer learning model and the reinforced data selector}
 \label{alg:joint-learning}
\end{algorithm}


\section{Experiments}
\label{sec:exp}
\subsection{Data Description}
\label{subsec:data-desc}

In this paper, we follow previous work~\cite{Yu2018TL} and use paraphrase identification and natural language inference data to evaluate the performance of our RTL model on text matching. Four benchmark datasets are used in the PI and NLI tasks. Both task settings are designed to transfer from a relatively open domain to a closed domain. Statistics for all datasets are presented in Table~\ref{tab:data-stat}.

\subsubsection{\textbf{Natural Language Inference (NLI)}}
We use MultiNLI~\cite{MultiNLI} as the source domain data and SciTail~\cite{scitail} as the target domain data. \textbf{MultiNLI} is a large crowdsourced corpora for textual entailment recognition. Each sample is a (premise, hypothesis, label) triplet, where the label is one of the ENTAILMENT, NEUTRAL, and CONTRADICTION.
In contrast to another widely used NLI dataset SNLI~\cite{SNLI}, where all premise sentences are derived from image captions, MultiNLI has more diverse text sources and thus is more suitable to serve as the source domain in a TL setting. We use the 1.0 version of MultiNLI with the training data from all five domains. We discard the samples with no gold labels. \textbf{SciTail} is a recently released textual entailment dataset in the science domain. In contrast to SNLI and MultiNLI, the premises and hypotheses in SciTail are generated with no awareness of each other.
Therefore SciTail is more diverse in terms of linguistic variations and thus is more challenging than other entailment datasets~\cite{scitail}. However, the labels in SciTail only consists of ENTAILMENT and NEUTRAL. Therefore, we remove the CONTRADICTION samples from MultiNLI. 

\subsubsection{\textbf{Paraphrase Identification (PI)}}
We use the Quora Question Pairs\footnote{https://www.kaggle.com/c/quora-question-pairs} as the source domain data and a paraphrase dataset made available in CIKM AnalytiCup 2018\footnote{https://tianchi.aliyun.com/competition/introduction.htm?raceId=231661} as the target domain data. \textbf{Quora Question Pairs} (Quora QP) is a large paraphrase dataset released by Quora\footnote{https://www.quora.com/}. Quora is a knowledge sharing website where users post questions and write answers for other users' questions. Due to the large amount of visitors, the user-generated questions contains duplications. Thus Quora released this dataset to encourage the research on paraphrase identification. \textbf{AnalytiCup Data} consists of question pairs in the E-commerce domain. It was released with the CIKM AnalytiCup 2018. This competition targets the research problem of cross-lingual text matching. This dataset contains labeled English training data and unlabeled Spanish data. However, in this work, we only deal with the labeled English data. We sample the training, validation, and testing data for the AnalytiCup data since no pre-defined data partitions are available.

\begin{table}[htbp]
\footnotesize
\caption{Data Statistics. For source domains, only training data is used. The numbers before and after the ``/'' are \# all examples and \# positive examples. ``Positive'' refers to PARAPHRASE in PI and ENTAILMENT in NLI.}
\label{tab:data-stat}
\vspace{-0.3cm}
\begin{tabular}{@{}llllll@{}}

\toprule
Task                 & Domain & Data       & Train           & Validation & Test      \\ \midrule
\multirow{2}{*}{PI}  & Source & Quora QP   & 404,287/149,263 & N/A        & N/A       \\
                     & Target & AnalytiCup & 6,668/1,731     & 3,334/830  & 3,330/820 \\ \midrule
\multirow{2}{*}{NLI} & Source & MultiNLI   & 261,799/130,899 & N/A        & N/A       \\
                     & Target & SciTail    & 23,596/8,602    & 1,304/657  & 2,126/842 \\ \bottomrule
\end{tabular}
\end{table}

\subsection{Experimental Setup}
\label{subsec:exp-setup}

\subsubsection{\textbf{Baselines}}
\label{subsubsec:baselines}
We consider the following baselines:
\begin{itemize}[leftmargin=1em]
    \item \textbf{Base model}~\cite{DBLP:conf/emnlp/Parikh2016DAM}: we use the shared encoder described in Section \ref{subsec:base-model} with a classification layer to form a decomposable attention model. This base model is trained with the target domain data.
    \item \textbf{Transfer baseline}: we use the TL model described in Section \ref{subsec:transfer-learning-model} to provide a stronger baseline. The transfer baseline leverages training data in both source and target domains.
    \item \textbf{Ruder and Plank}~\cite{ruder2017select} proposed a data selection method with Bayesian optimization for transfer learning. This data selection approach is model-independent. We use it on the top of our transfer learning model to keep the comparisons fair.
\end{itemize}

\subsubsection{\textbf{Evaluation Metrics}}
\label{subsubsec:metrics}
For both tasks, we adopt accuracy (Acc) and the Area under the ROC curve (AUC) as evaluation metrics. Significance tests can only be performed on accuracy.

\subsubsection{\textbf{Implementation Details}}
\label{subsubsec:parameter}
We present the parameter settings and implementation details as follows. All models are implemented with TensorFlow\footnote{\url{https://www.tensorflow.org/}}. Size for the hidden layers of the decomposable attention model is 200. The max sequence length is 40 for PI and 50 for NLI. The padding is masked to avoid affecting the gradient. Hyper-parameters including the size of the hidden layer of the policy network, and the reward discount factor are tuned with the target domain validation data. Checkpoints are saved at the end of every epoch and produce an evaluation on the test set. All models are trained with a NVIDIA Titan X GPU using Adam~\cite{DBLP:journals/corr/KingmaB14}. The initial learning rate is 0.001 for the transfer model and 0.02 for the policy network. The parameters of Adam, $\beta_1$ and $\beta_2$ are 0.9 and 0.999 respectively. The hidden layer size and optimization methods for the value estimator network are the same with the policy network. 

The transfer learning model is pre-trained for 50 iterations for both tasks before the reinforced data selector is applied. For the word embedding layer, we use GloVe~\cite{pennington2014glove} (840B tokens) to initialize the embedding look up table. The dimension of word embedding is 300. Word vectors are set to trainable.

\subsection{Evaluation Results}
\label{subsec:evaluation-results}
We present the evaluation results in Table~\ref{tab:main-results}. Models are tuned with the target domain validation data and results are reported on the target domain testing data. 

\begin{table}[htbp]

\caption{Testing performance in the target domain for PI and NLI tasks. Our model is referred to as RTL. The significance tests can only be performed on accuracy. $\ddagger$ means statistically significant difference over the strongest baseline with $p < 0.01$ measured by the Student's paired t-test.}
\label{tab:main-results}
\vspace{-0.3cm}
\begin{tabular}{@{}lllll@{}}
\toprule
\multirow{2}{*}{Methods}           & \multicolumn{2}{c}{PI} & \multicolumn{2}{c}{NLI} \\ \cmidrule(l){2-3} \cmidrule(l){4-5} 
                                   & Acc        & AUC       & Acc        & AUC        \\ \midrule
Base Model~\cite{DBLP:conf/emnlp/Parikh2016DAM}                         & 0.8393     & 0.8548    & 0.7300     & 0.7663     \\ 
Transfer Learning Model            & 0.8488     & 0.8706    & 0.7453     & 0.8044     \\ 
Ruder and Plank~\cite{ruder2017select}                    & 0.8458     & 0.8680    & 0.7521     & 0.8062     \\ 
RTL & \textbf{0.8616}$^\ddagger$     & \textbf{0.8829}    & \textbf{0.7672}$^\ddagger$     & \textbf{0.8163}     \\ \bottomrule
\end{tabular}
\end{table}

For paraphrase identification, we observe that the base model alone achieves relatively good performance. The TL model manages to have a minor improvement over the base model. However, the data selection method with Bayesian optimization by Ruder and Plank~\cite{ruder2017select} fails to make further improvement over the TL model. Based on the base model performance, we speculate that the PI task on AnalytiCup data is a relatively straightforward task. Therefore, it could be possible that sophisticated models do not always boost the performance. Under these circumstances, our model (RTL) still manages to generate a statistically significant improvement over the strongest baseline.

For natural language inference, the performance on all models are lower in general compared to the PI task. This is due to the fact that SciTail is very challenging as described in Section~\ref{subsec:data-desc}. The base model has moderate performance on this task. The TL model improves the performance thanks to the source domain data. The data selection method with Bayesian optimization manages to make a further improvement, indicating the large potential of data selection in this setting. Moreover, our RTL model outperforms the strongest baseline by a large margin with statistical significance. These results demonstrate the effectiveness of the RTL model. 

In Ruder and Plank~\cite{ruder2017select}, the TL model is considered as a sub-module of the Bayesian optimization based data selection framework. This framework evaluates the utility of the data selection based on the final performance of the TL model. In our RTL framework, the TL model and the reinforced data selector are trained jointly and thus the data selection policy is updated more efficiently and effectively. This could be the reason behind the improvement of our model over Ruder and Plank~\cite{ruder2017select}.

\subsection{Ablation Analysis}
\label{subsec:ablation-analysis}
In addition to the best performing model in Section~\ref{subsec:evaluation-results}, we also investigate different variations of the RTL model. The variations are made in terms of three aspects: the reward functions, optimization methods for the policy network, and state representations. 

\subsubsection{\textbf{Reward Functions and Policy Optimization Methods.}} We consider various reward functions and policy optimization methods as the main settings for our ablation tests. The results are presented in Table~\ref{tab:ablation-results}.

\begin{table}[htbp]
\caption{Testing performance of RTL with different variations. The last entry is the final RTL model in Table~\ref{tab:main-results}.}
\label{tab:ablation-results}
\begin{tabular}{llllll}
\toprule
\multicolumn{2}{c}{Methods} & \multicolumn{2}{c}{PI} & \multicolumn{2}{c}{NLI} \\  \cmidrule(l){1-2} \cmidrule(l){3-4}  \cmidrule(l){5-6} 
Reward   & RL  & Acc        & AUC       & Acc        & AUC        \\ \midrule
AUC      & REINFORCE        & 0.8557     & 0.8818    & 0.7486     & 0.8070     \\
AUC      & Actor-Critic     & 0.8545     & 0.8793    & 0.7613     & 0.8067     \\
Acc      & REINFORCE        & 0.8428     & 0.8788    & 0.7587     & 0.8121     \\
Acc      & Actor-Critic     & \textbf{0.8616}     & \textbf{0.8829}    & \textbf{0.7672}     & \textbf{0.8163}     \\ \bottomrule
\end{tabular}
\end{table}

As shown in Table~\ref{tab:ablation-results}, we experiment with two reward functions of using accuracy or AUC. Also, we use two algorithms to optimize the policy network: the REINFORCE algorithm and the actor-critic algorithm. We observe that policy networks optimized with the actor-arctic algorithm generally produce similar or better performance. On the other hand, when using the same algorithm to optimize the policy network, using accuracy as the reward tends to generate better results. The best setting is to use accuracy as the reward and actor-critic for policy optimization. Thus, we adopt this setting for our final RTL model. 

\subsubsection{\textbf{State Features.}} In addition to the model variations on main settings of reward functions and policy optimization methods, we also perform a state feature ablation test under the best main setting. We have five state features in total as mentioned in Section~\ref{subsubsec:RL-state}. Four of them (feature 2, 3, 4, 5) can be considered as a feature group because they are all designed to evaluate whether a source sample can be easily classified by the TL model. Thus, we perform a feature ablation test on two feature groups.

\begin{table}[htbp]
\caption{Testing performance of RTL with different state features under the best main setting of (Acc, Actor-Critic). The TL model without data selection is also included for comparisons. The last entry is the final RTL model in Table~\ref{tab:main-results}.}
\label{tab:feature-ablation-results}
\vspace{-0.3cm}
\begin{tabular}{@{}lllll@{}}
\toprule
\multicolumn{1}{c}{\multirow{2}{*}{Features}} & \multicolumn{2}{c}{PI} & \multicolumn{2}{c}{NLI} \\ \cmidrule(l){2-3}  \cmidrule(l){4-5} 
                         & Acc        & AUC       & Acc        & AUC        \\ \midrule
Transfer Learning Model  & 0.8488     & 0.8706    & 0.7521     & 0.8044          \\              
(1)                      & 0.8539     & 0.8813    & 0.7594     & 0.8135     \\
(2) (3) (4) (5)          & 0.8529     & 0.8778    & 0.7507     & 0.7916     \\
(1) (2) (3) (4) (5)      & \textbf{0.8616}     & \textbf{0.8829}    & \textbf{0.7672}     & \textbf{0.8163}     \\ \bottomrule
\end{tabular}
\end{table}

As shown in Table~\ref{tab:feature-ablation-results}, the reinforced data selector with the second feature group (feature 2 - feature 5) achieves similar or higher results than the transfer baseline. This indicates that hand-crafted features in the second feature group have limited capacity for the state representation. On the other hand, feature 1 achieves relatively good performance when used alone. This suggests that the hidden representations of source samples learned by the shared encoder are capable of providing good descriptions of the model's states. Moreover, the model gives the best performance if we combine the two feature groups, confirming that all features contribute to the model performance.

\subsection{Impact of Hyper-parameters}
\label{subsec:hyper-params}
\begin{figure}[ht]
	\centering
	\begin{subfigure}[b]{0.253\textwidth}
		\includegraphics[width=\textwidth]{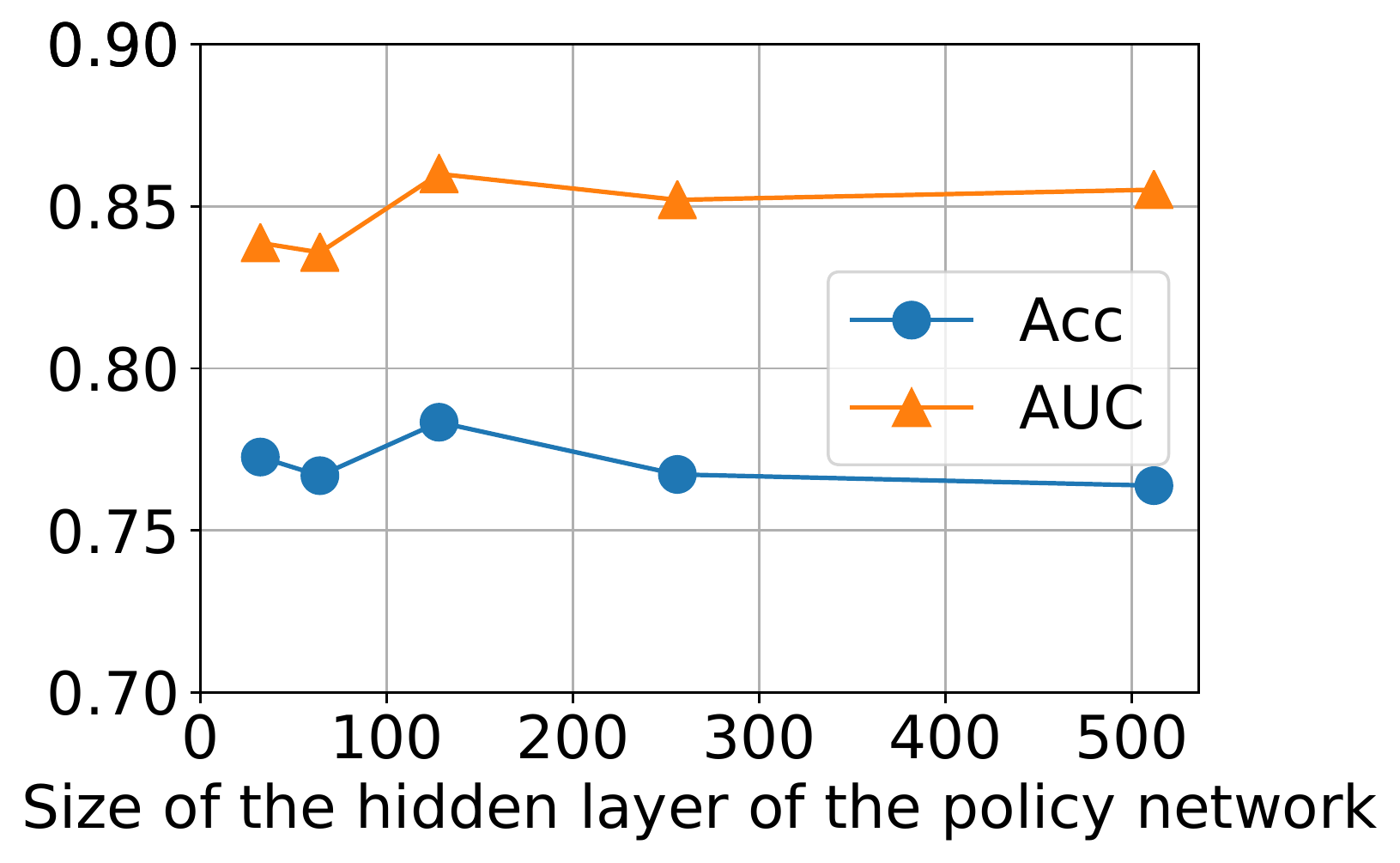}
		\label{fig:nb-policy}
	\end{subfigure}
	\begin{subfigure}[b]{0.217\textwidth}
        \includegraphics[width=\textwidth]{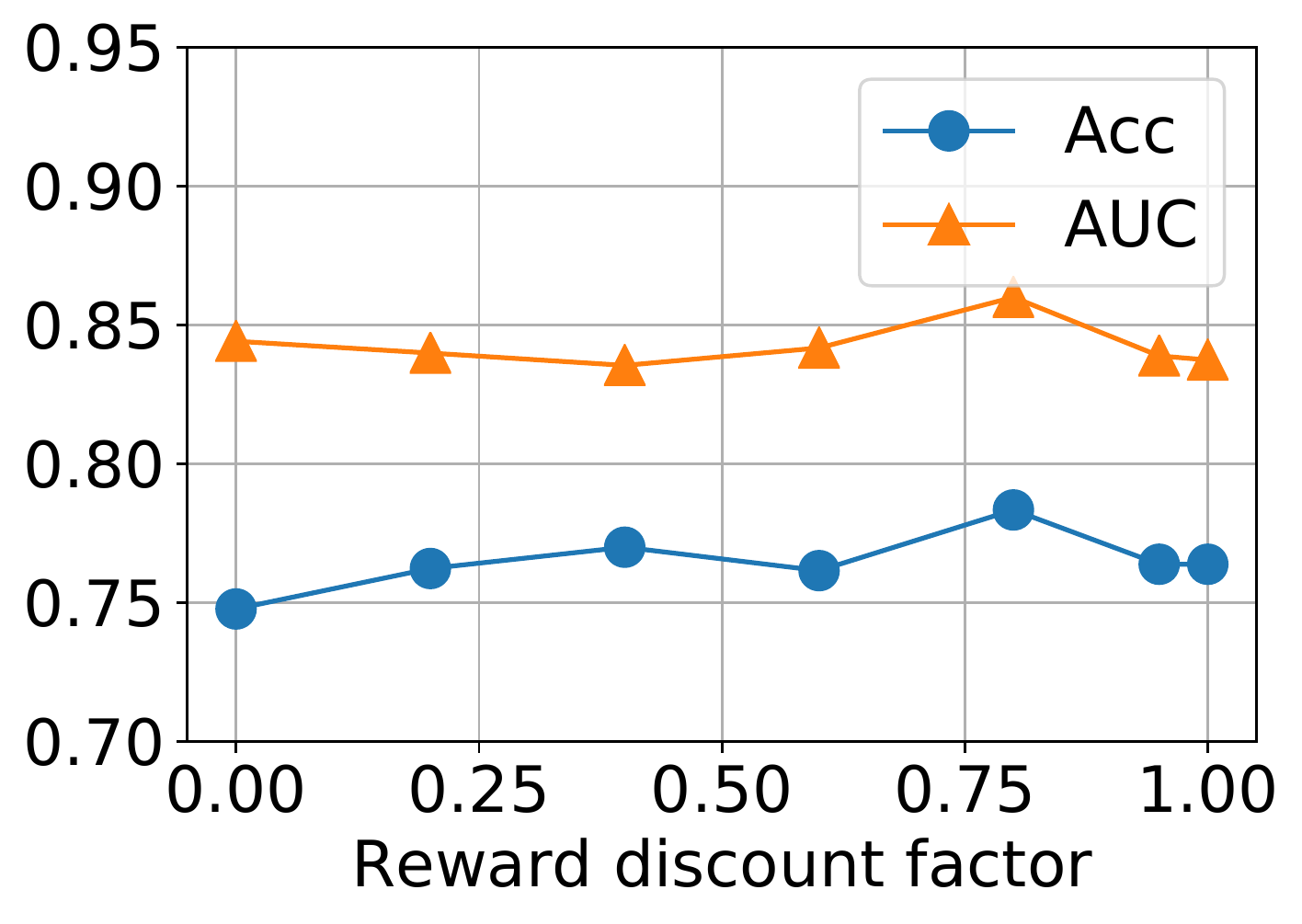}
		\label{fig:reward-decay}
	\end{subfigure}
	\vspace{-1cm}
	\caption{Impact of hyper-parameters of the reinforced data selector on the validation data of the NLI task.}
	\label{fig:tune}
\end{figure}

We use the NLI task to demonstrate the impact of two hyper-parameters on our model: the size of the hidden layer of the policy network and the reward discount factor. Both hyper-parameters are related to the reinforced data selector. The number of units in the hidden layer of the policy network is tuned in (32, 64, 128, 256, 512). The choices for the reward discount factor are (0, 0.2, 0.4, 0.6, 0.8, 0.95, 1). The reward discount factor = 0 denotes that future rewards are not considered when updating the policy, while reward discount = 1 means to fully consider future rewards without any discount. Figure~\ref{fig:tune} presents the validation performance with different hyper-parameters. The performance of the transfer model has a peak value when the hidden layer of the policy network has 128 units. This indicates that too small or too large capacity of the policy network cannot benefit the data selection process. In addition, the model performance does not seem be heavily influenced by the different small reward discount factors. The trend suggests that the reinforced data selector can benefit more when contributions of previous actions are properly emphasized with relatively large reward discount factors but not too large.

\vspace{-0.2cm}

\subsection{Case Study and Performance Interpretation}
\label{subsec:case-study}
The results in Section~\ref{subsec:evaluation-results} demonstrate the effectiveness of our method. However, due to the lack of interpretability of neural architectures, it is difficult to speculate the reasons behind the decisions made by the reinforced data selector. Therefore, instead of trying to interpret specific cases, we present an overall performance interpretation of our method by measuring the distance between domains of interest. Specifically, we compute the Wasserstein distance between the term distributions of the target domain and the source domains. The source domains include the source domain data selected or dropped by the reinforced data selector and the randomly selected source domain data.

Wasserstein distance is also known as the earth mover's distance. It measures the distance between two probability distributions on a given metric space $\mathbb{R}$. Intuitively, it can be considered as the minimum amount of work required to transform the distribution $u$ to the distribution $v$, which can be computed by the amount of earth it has to be moved, multiplied by the distance it has to be moved. Formally, the 1\textsuperscript{st} Wasserstein distance is defined as:
\begin{equation}\label{eqn:wasserstein-distance}
\begin{aligned}
W_1(u, v) = \inf_{\pi \in \Gamma(u, v)} \int_{\mathbb{R}\times\mathbb{R}}\!|x-y|\,\text{d}\pi(x, y)
\end{aligned}
\end{equation}
where $\Gamma(u, v)$ is a set of probability distributions on $\mathbb{R}\times\mathbb{R}$ and $u, v$ are two probability distributions. In our case, $u, v$ are defined as the term frequency distributions on two domains respectively. 

Besides, Wasserstein distance can handle the distributions where some events have the probability of $0$ (a certain word presents in one domain but not in the other domain). Also, Wasserstein distance is a symmetric measure, meaning that $W_1(u, v)$ is equal to $W_1(v, u)$. These properties make it suitable for our task.

We keep track of the actions taken by the reinforced data selector. The selected data at the final episode is considered as the final selected data. We compute the Wasserstein distance between the target domain data and the source domain data, including the selected and dropped source domain data. We also include the randomly selected source domain data for comparison. The number of randomly selected instances is identical to the number of instances selected by the reinforced data selector. The results are presented in Table~\ref{tab:wasserstein-distance}. Due to the large amount of tokens in the source domain data, the normalized term frequency for any given term is relatively small, and thus the the Wasserstein distance is small in terms of the order of magnitude.

\begin{table}[htbp]
\caption{The Wasserstein distances between the term distributions of different domains.}
\label{tab:wasserstein-distance}
\vspace{-0.3cm}
\begin{tabular}{@{}llll@{}}
\toprule
Name & Domains in Comparison    & PI        & NLI       \\ \midrule
$D_{origin}$  & Target $\leftrightarrow$ Source           & 5.250E-06 & 3.256E-06 \\
$D_{select}$  & Target $\leftrightarrow$ Source (Selected) & 4.963E-06 & 3.190E-06 \\
$D_{drop}$   & Target $\leftrightarrow$ Source (Dropped)  & 5.320E-06 & 3.290E-06 \\
$D_{rand}$    & Target $\leftrightarrow$ Source (Random)   & 5.232E-06 & 3.243E-06 \\ \bottomrule
\end{tabular}
\end{table}

We observe the exact same patterns for the PI and NLI tasks: (1) $D_{rand} \approx D_{origin}$, which means that random selection only influences the term distribution slightly. This sets a baseline for other distances. (2) $D_{select} < D_{origin}$, which means that the source domain data selected by the reinforced data selector is closer to the target domain data and thus may contribute to the transfer learning process. (3) $D_{drop} > D_{origin}$, which means that the source domain data dropped by the reinforced data selector is not very similar to the target domain data and thus may cause negative transfer. These findings indicate that our method is able to select source domain data whose Wasserstein distance is close to the target domain data. This is reasonable as such source domain data could be more transferrable and helpful for the target domain. 

\section{Conclusions and Future work} 
\label{sec:conclusion}

In this paper, we proposed a reinforced data selection method in a DNN based transfer learning setting. Specifically, we used reinforcement learning to train a data selection policy to select high-quality source domain data with the purpose of preventing negative transfer. We investigated different settings of states, rewards, and policy optimization methods to test the robustness of our model. Extensive experiments on PI and NLI tasks indicate that our model can outperform existing methods with statistically significant improvements. Finally, we used Wasserstein distance to measure the source and target domain distances before and after the data selection. This study indicates that our method is capable of selecting source domain data that has a similar probability distribution to the target domain data. Future work will consider to explore more effective state representations and adapt our method to other tasks.


\begin{acks}
This work was supported in part by the Center for Intelligent Information Retrieval. Any opinions, findings and conclusions or recommendations expressed in this material are those of the authors and do not necessarily
reflect those of the sponsor.

\end{acks}

\bibliographystyle{ACM-Reference-Format}
\bibliography{acmart} 

\end{document}